\documentclass[a4paper,twocolumn,showpacs,superscriptaddress,floatfix,nofootinbib]{revtex4-1}
\usepackage{graphicx}
\usepackage{epsf}
\usepackage{epsfig}
\usepackage{amssymb,amsmath}
\usepackage[usenames]{color}
\usepackage{amssymb}
\usepackage{times}
\usepackage{soul}
\usepackage{hyperref}
\hypersetup{
  colorlinks=true,        
  linkcolor=blue,         
  citecolor=cyan,         
}

\newcommand{\cf}{cf.~}
\newcommand{\ie}{i.e.,~}
\newcommand{\eg}{e.g.,~}

\renewcommand{\BibitemShut}[1]{}
\begin{document}

\title{Did GW150914 produce a rotating gravastar?}
\author{Cecilia Chirenti} 
\affiliation{Centro de Matem\'atica, Computa\c
  c\~ao e Cogni\c c\~ao, UFABC, 09210-170 Santo Andr\'e-SP, Brazil}
\author{Luciano~Rezzolla}
\affiliation{Institut f{\"u}r Theoretische Physik,
  Max-von-Laue-Stra{\ss}e 1, 60438 Frankfurt, Germany}
\affiliation{Frankfurt Institute for Advanced Studies, Ruth-Moufang-Str. 1,
60438 Frankfurt, Germany}

\begin{abstract}
The interferometric LIGO detectors have recently measured the first
direct gravitational-wave signal from what has been interpreted as the
inspiral, merger and ringdown of a binary system of black holes. The
signal-to-noise ratio of the measured signal is large enough to leave
little doubt that it does refer to the inspiral of two massive and
ultracompact objects, whose merger yields a rotating black hole. Yet, the
quality of the data is such that some room is left for alternative
interpretations that do not involve black holes, but other objects that,
within classical general relativity, can be equally massive and compact,
namely, gravastars. We here consider the hypothesis that the merging
objects were indeed gravastars and explore whether the merged object
could therefore be not a black hole but a rotating gravastar. After
comparing the real and imaginary parts of the ringdown signal of GW150914
with the corresponding quantities for a variety of gravastars, and
notwithstanding the very limited knowledge of the perturbative response
of rotating gravastars, we conclude it is not possible to model the
measured ringdown of GW150914 as due to a rotating gravastar.
\end{abstract}


\maketitle


\emph{Introduction.~~} Gravastars were proposed in 2004 by Mazur and
Mottola \cite{Mazur2004} as an ingenious alternative to the end state of
stellar evolution for very massive stars, that is, as an alternative to
black holes. The name gravastar comes from ``gravitational vacuum
condensate star" and it was proposed to be almost as compact as a black
hole, but without an event horizon or a central singularity. This object
would be formed as gravitational collapse brought the stellar radius very
close to its Schwarzschild radius and as a phase transition would form a
de Sitter core. This ``repulsive'' core stabilises the collapse, while
the baryonic mass ends as a shell of stiff matter surrounding the
core. Despite their uncertain and rather exotic origin, gravastars are
perfectly acceptable solutions of the Einstein equations within classical
general relativity.

Considerable effort has been dedicated to study gravastars, for instance
exploring different possibilities for its structure \cite{Visser2004,
  Cattoen2005}, generalising the solution \cite{Horvat2008, Turimov2009,
  Uchikata2015} and investigating possible observational signatures
\cite{Broderick2007, Sakai2014, Pani2015}. As alternatives almost
indistinguishable from a black hole in terms of electromagnetic
radiation, gravastars have attracted the attention of those who wished
for a spacetime solution without the issues brought by the existence of
singularities and event horizons. Work was also done in order to assess
its viability, in particular looking for instabilities in the solutions.
Hence, there have been studies on the stability against radial
oscillations \cite{Visser2004, Horvat2011} and axial and polar
gravitational perturbations \cite{DeBenedictis2005, chirenti_2007_htg,
  Pani2009}. For slowly rotating gravastars, scalar perturbations in the
context of the ergoregion instability were also studied
\cite{Chirenti2008, Cardoso2007}. None of these works has pointed out to
a response that would allow one to discard gravastars as plausible
solutions of general relativity.

In their original model, Mazur and Mottola \cite{Mazur2004} proposed a
gravastar with infinitesimal but nonzero thickness as this allowed them
to derive the most salient properties of the model analytically
\cite{Rezzolla_book:2013} (the thickness of the shell is
effectively zero in the gravastar model of \cite{Visser2004}). In any
astrophysically realistic configuration, however, the gravastar is
expected to have a finite thickness, so that the parameter space for
nonorotating gravastar solutions has effectively three degrees of
freedom: the total (gravitational) mass $M$, the inner radius of the
shell $r_1$, and the outer one $r_2$, which is also the radius of the
gravastar \cite{Visser2004, Cattoen2005, chirenti_2007_htg}. As a result,
gravastar solutions are normally classified in terms of $M$, of the
compactness $\mu \equiv M/r_2$, and of the thickness of the shell $\delta
\equiv r_2 -r_1$; within this parameter space it is always possible to
find stable solutions. Because of the freedom in choosing $\delta$, it is
in principle possible to build gravastars with $\delta /M \lll 1$ and
radius that is only infinitesimally larger than the Schwarzschild radius,
thus making these objects frustratingly hard to distinguish from black
holes when using electromagnetic emission. However, as pointed out almost
a decade ago \cite{chirenti_2007_htg}, it is possible to distinguish a
gravastar from a black hole if sufficiently strong gravitational
radiation is detected; with the recent observation of GW150914
\cite{Abbott2016a}, we can now start to do so.

In this Letter we consider the possibility that the event GW150914 was
produced by the merger of two gravastars, creating then a more massive
and rotating gravastar as a result of the merger. Given the strength of
the detected signal, the inspiral part of the signal could be reproduced
by two compact objects that are not necessarily black holes, and could
well be very compact gravastars or other exotic compact
objects.\footnote{The compactness argument presented in
  \cite{Abbott2016c} shows that the inspiralling bodies in GW150914 must
  have had radii smaller than 175 km, otherwise they would have touched
  before reaching the observed gravitational wave frequency of 150 Hz,
  therefore placing a lower bound of $\mu > G/c^2 (35\,
  M_{\odot})/(175\,{\rm km}) \sim 0.3$ on the compactness.} The ringdown,
however, presents a characteristic signature of the final compact
object. After performing a quasi-normal mode analysis of slowly rotating
gravastars, based on the rotational corrections of the oscillation
frequencies of compact stars, we have investigated whether the resulting
object from the binary merger in GW150914 could be a gravastar. We show
here that, within the possible accuracy of our results, that object
ringing down in GW150914 could not be a gravastar.

\medskip\emph{Results.~~} In its essential simplicity, the merger of a
black-hole binary system will lead to the formation of a new, rotating
black hole \cite{Pretorius:2005gq, Campanelli:2005dd, Baker:2005vv} (a
Schwarzschild black hole can in principle also be produced in a binary
black-hole merger \cite{Rezzolla-etal-2007b}, but this is rather
unlikely). The new Kerr black hole will be initially highly perturbed and
hence emit gravitational waves by oscillating in its quasi-normal modes
(QNMs); this is the characteristic ringdown signal of a perturbed black
hole. A large bulk of work has been produced to extract information from
the ringdown produced by the merger, including the main properties of the
newly formed black hole (\eg mass and spin) \cite{Berti:2007snb,
  Berti06b}, but also the recoil direction and magnitude
\cite{Schnittman:2007ij, Koppitz-etal-2007aa, Pollney:2007ss,
  London2014}.

The real and imaginary parts, $\sigma_r$ and $\sigma_i$, of the lowest
order $\ell=2=m$ QNM of a Kerr black hole have been studied in great
detail by a number of authors within perturbation theory already in the
90's (see, \eg Ref. \cite{Detweiler1980} for one of the initial analyses
and Ref. \cite{Kokkotas99a} for a comprehensive review). In
  particular, for a rotating black hole with mass $M$, angular momentum
  $J$ and dimensionless spin parameter $a \equiv J/M^2$, these
  frequencies have been shown to be very well approximated by simple
  expressions \cite{Berti06b}. Using perturbative analyses and
numerical-relativity simulations, the ringdown signal from the
observations of GW150914 was associated to the lowest QNM of a Kerr
black hole with dimensionless spin $a=0.68^{+0.05}_{-0.06}$ and mass $M =
62.2^{+3.7}_{-3.4}\,M_{\odot}$ \cite{Abbott2016a,Abbott2016d}.

\begin{figure}
\centering
\includegraphics[width=0.80\columnwidth]{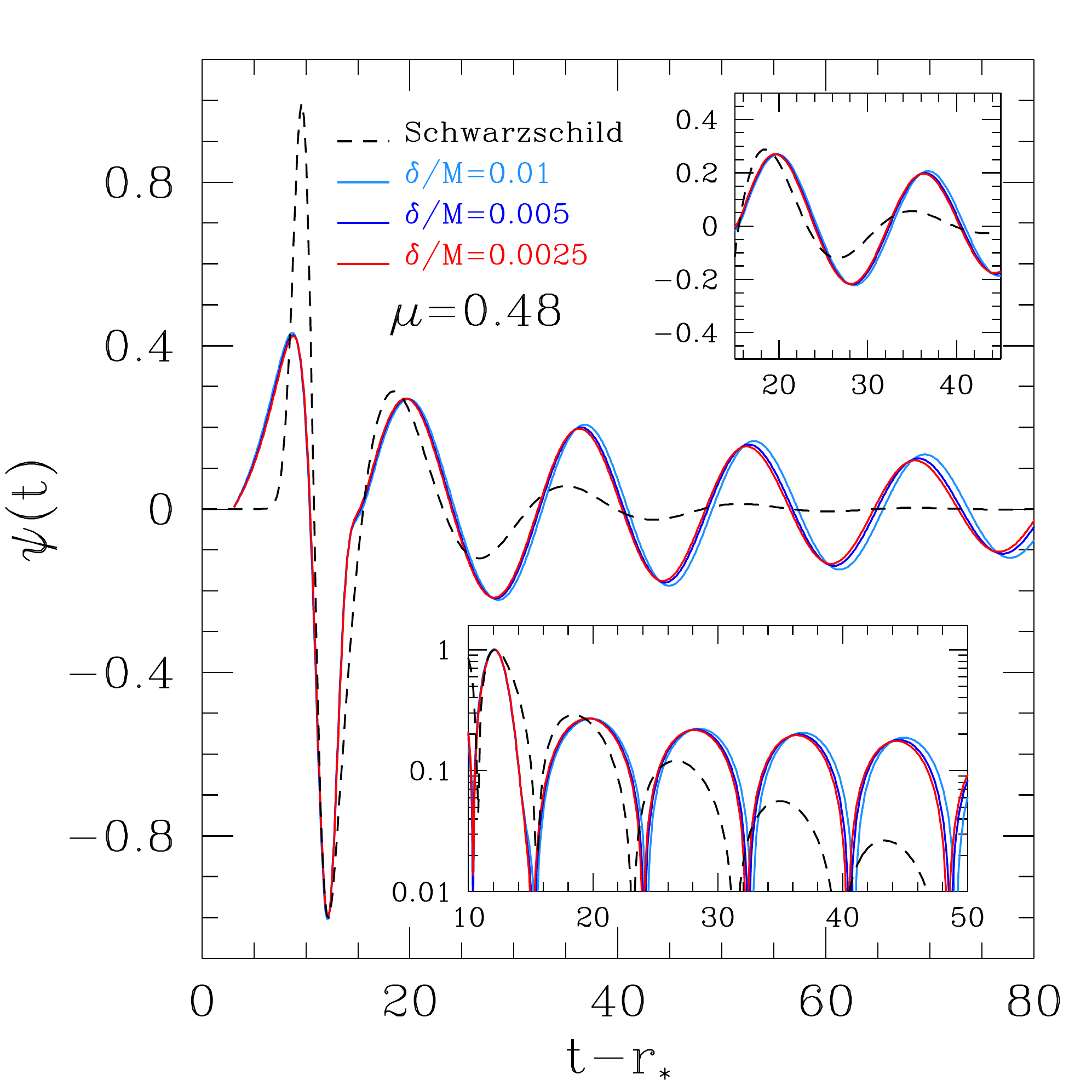}
\caption{Evolution of the $\ell=2=m$ axial perturbation $\psi(t)$ for a
  Schwarzschild black hole (black dashed line) and for three different
  gravastars (coloured solid lines) with decreasing thickness,
  $\delta/M=0.01, 0.005$ and $0.0025$; in all cases the gravastars have
  the same compactness $\mu=0.48$ and the time is retarded with a
  tortoise coordinate $r_*$ \cite{chirenti_2007_htg}. Note that even for
  $\delta/M=0.0025$ the QNMs can be distinguished clearly over a
  dynamical timescale.}
\label{fig:f1} 
\end{figure}

The vast literature on the perturbations of rotating black holes is in
stark contrast with the very limited knowledge of perturbed rotating
gravastars. Indeed, essentially all of the work carried out so far on the
QNMs of gravastars has concentrated on nonrotating models and
gravitational perturbations (axial and polar) \cite{DeBenedictis2005,
  chirenti_2007_htg, Pani2009}. An example of the perturbative response of
a gravastar is reported in Fig. \ref{fig:f1}, which shows the evolution
of the $\ell=2=m$ axial perturbation $\psi(t)$ for a Schwarzschild black
hole (black dashed line) and for three different gravastars (coloured
solid lines) with decreasing thickness, \ie with $\delta/M=0.01, 0.005$
and $0.0025$; in all cases the gravastars have the same compactness
$\mu=0.48$.

Figure \ref{fig:f1} is also useful to clarify a point that may
  otherwise be a source of confusion. Ref.~ \cite{Cardoso2016} has
  recently discussed that an early-time decay of a gravitational-wave
  signal induced by a perturbation of an ultra-compact object is similar
  to that of a black hole, irrespective of the differences in the QNM
  spectrum. We obviously agree with this conclusion but also note that
  whether or not one is able to distinguish the very early-time signal
  (assuming this is all that is observed) depends on how close the
  surface of the ``black-hole mimicker'' is to the event horizon. Figure
  \ref{fig:f1} shows that even in the extreme case of very thin and
  ultracompact gravastars with $\delta/M=0.0025,\ \mu=0.48$, whose
  surface is only $4\%$ (in radius) outside of the event horizon, even
  the very early part of the ringdown can be distinguished from the
  corresponding one for a Schwarzschild black hole, while the late part
  of the ringdown will be considerably different (as it should be in
  order to yield different QNMs). This conclusion holds true for any
  realistic gravastar independently of the internal structure and as long
  as the surface is not at an \emph{infinitesimal} distance away from the
  putative horizon position $2M$ (see also the discussion in
  Ref. \cite{Konoplya2016b}). This is because at the gravastar's
surface, where the fields are strong and dynamical, the two spacetimes
will be different, as will be boundary conditions of the
  corresponding perturbative problem.  The timescale over which this
difference can be probed via a perturbation is given by the crossing time
between the peak of the scattering potential and the surface of the star;
this timescale is comparable but smaller than the dynamical timescale of
the gravastar's response and so the difference is clearly detectable on
such a timescale (\cf Fig. \ref{fig:f1}).

\begin{figure}
\centering
\includegraphics[width=0.80\columnwidth]{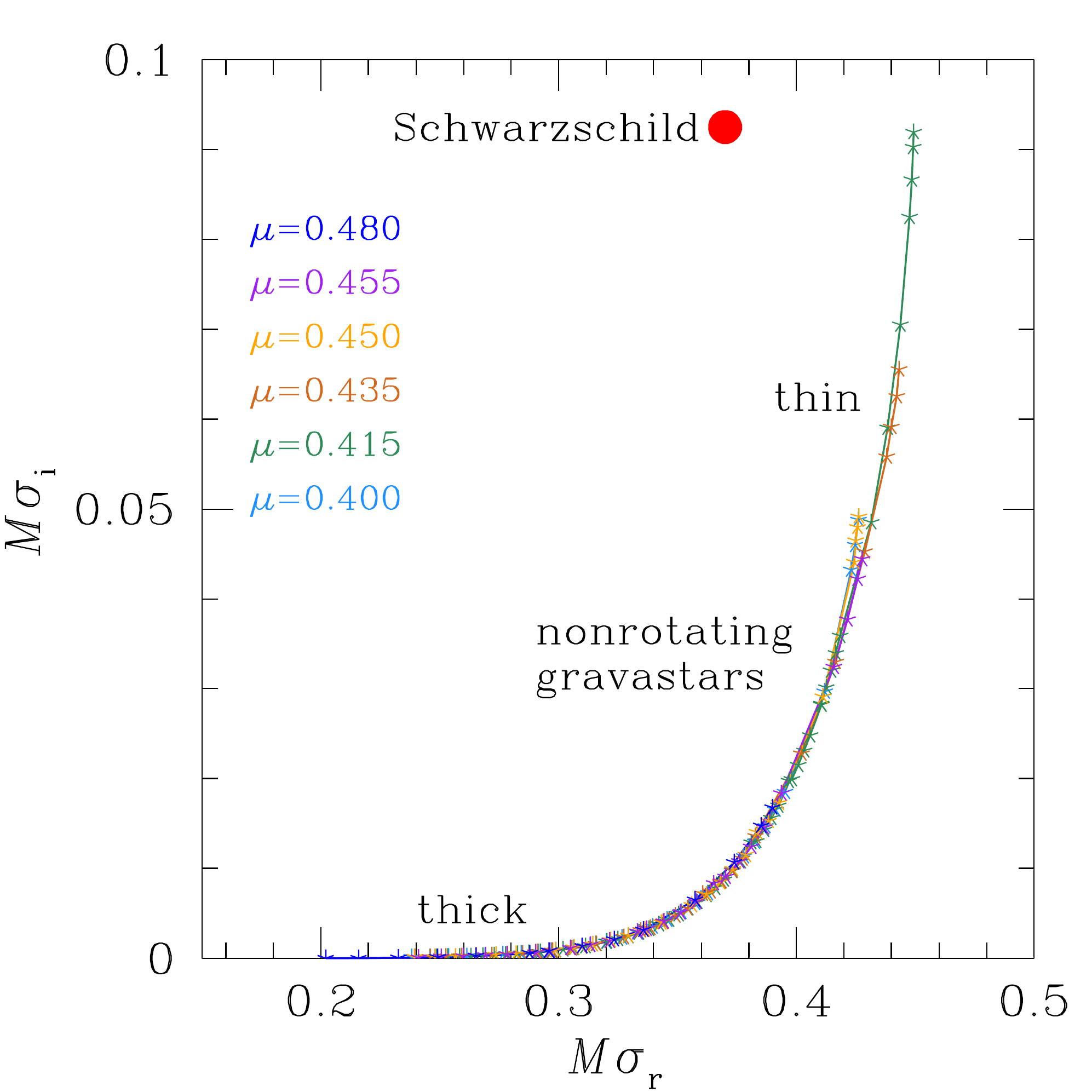}
\caption{Real and imaginary parts, $\sigma_r, \sigma_i$, of the
  eigenfrequencies as computed for the $\ell=2=m$ axial QNM of
  nonrotating gravastars \cite{chirenti_2007_htg}. Lines of different
  colours refer to sequences of gravastars with constant compactness
  $\mu$, but varying thickness $\delta$. Thick gravastars have small
  eigenfrequencies which increase as the gravastars become increasingly
  thin. Note however that the eigenfrequencies remain distinct from that
  of a Schwarzschild black hole (solid red circle).}
\label{fig:f2}
\end{figure}

The real and imaginary parts of the $\ell=2=m$ eigenfrequencies for axial
gravitational perturbations are reported in Fig. \ref{fig:f2}, for a
variety of nonrotating gravastars \cite{chirenti_2007_htg}. Lines of
different colours refer to sequences of gravastars with constant
compactness $\mu$, but varying thickness $\delta/r_2$, which is marked by
the various points on the curves and that ranges from $\delta/r_2=5\times
10^{-4}$ to $\delta/r_2=0.3$. The various $\mu={\rm const.}$ curves
essentially overlap in the $(\sigma_r, \sigma_i)$ plane, with ``thick''
gravastars yielding low oscillation frequencies and long damping times,
while ``thin'' gravastars show instead high oscillation frequencies and
short damping times. Such a behavior is rather natural since thin
(nonrotating) gravastars tend to be increasingly similar to a
(nonrotating) black hole, whose eigenfrequencies are shown with a red
circle. Yet, independently of the compactness and thickness considered,
the eigenfrequencies of nonrotating gravastars differ from the
corresponding eigenfrequencies of a nonrotating black hole
\cite{chirenti_2007_htg}.

The analysis of the QNMs of rotating gravastars is effectively limited to
scalar perturbations in the context of the ergoregion instability
\cite{Kokkotas2004, Chirenti2008, Cardoso2007}. To make some progress
despite the scarce present knowledge we have exploited the considerable
experience that has been built in modelling the eigenfrequencies of
rotating stars from the knowledge of the (fundamental) $f$-mode
eigenfrequencies for nonrotating stars \cite{Kojima1993, Ferrari2007,
  Doneva:2013}. Interestingly, this seems to be a rather successful route
and useful results have been obtained using different techniques and
approximations. Because we are here interested in both the real and
imaginary parts of the eigenfrequencies, we have followed the work of
Ref. \cite{Ferrari2007}, who have shown that these frequencies can be
approximated as
\begin{align}
\label{eq:sigma_1}
\sigma_r &\simeq \sigma_{r,0}
\left(1 + m\, \epsilon\, \sigma'_r \right) + \mathcal{O}(\epsilon^2)\,,& \\
\label{eq:sigma_2}
\sigma_i &\simeq \sigma_{i,0}
\left(1 + m\, \epsilon\, \sigma'_i \right) + \mathcal{O}(\epsilon^2)\,,& 
\end{align}
where $\sigma_{r,0} (\sigma_{i,0})$ are the real (imaginary) parts of the
$f$ mode eigenfrequencies for the corresponding nonrotating star.

Expressions \eqref{eq:sigma_1}$-$\eqref{eq:sigma_2} contain two
corrections to the nonrotating eigenfrequencies, namely, $\epsilon$ and
$\sigma'_{r,i}$. The first one accounts for the corrections due to
rotation and is therefore proportional to the angular frequency of the
star $\Omega$ as measured when normalised to the maximum rotation
frequency, \ie the Keplerian frequency $\Omega_{_{\rm K}}$. The latter is
in general a complex function of the stellar structure, but is quite
robustly related to the average ``density'' of the star, as $\sqrt{\langle
\rho \rangle} \sim \sqrt{M_0/R^3} \sim \sqrt{M/R^3}$, where $M_0$ and $M$
are the rest-mass and gravitational mass of the star. As customary, we
express $\epsilon$ as
\begin{equation}
\epsilon \equiv \frac{\Omega}{\Omega_{_{\rm K}}} 
\simeq \frac{J/(MR^2)}{\Omega_{_{\rm K}}}
\simeq \chi \sqrt{\mu}
\,,
\end{equation}
where $J$ is the angular momentum of the star and $\chi \equiv
J/M^2$.

The second corrections in \eqref{eq:sigma_1}$-$\eqref{eq:sigma_2} are
instead given by the modifications in the eigenfrequencies due to changes
in compactness, \ie, $\sigma'_{r,i} = \sigma'_{r,i}(\mu)$. In principle,
these corrections should be obtained after performing a complete
perturbative analysis of rotating gravastars, in analogy with what has
already been done for relativistic stars \cite{Ferrari2007,
  Doneva:2013}. Because so little is known about the perturbative
response of rotating gravastars, we exploit all of the understanding of
the perturbative response of rotating compact stars to make progress in
the phase space of rotating gravastars.

Given the extreme equation of state and the even more bizarre de Sitter
interior, it is natural to ask how closely does the perturbative response
of a gravastar resemble that of a compact star. Answering this question
is not simple since gravastars have the thickness as an additional degree
of freedom given when compared to compact stars. However, what is
relevant here is that gravastars are essentially ultracompact stars with
a de-Sitter core exhibiting a behaviour that correlates closely with the
global properties such as the mass, the compactness or the effective
average densities \cite{chirenti_2007_htg}. Such correlations are not
surprising since gravastars ultimately have trapping potentials that are
very similar to those already encountered in ordinary ultracompact stars,
where QNMs can be trapped \cite{Chandra1991, Kokkotas2004} (\cf left
panel of Fig. 6 in Ref. \cite{chirenti_2007_htg}).

Since the similarities discussed above suggest that it is not
unreasonable to use knowledge on compact stars also in the context of
gravastars, and lacking any alternative approach at present, we have used
the expressions for $\sigma'_{r,i} = \sigma'_{r,i}(\mu)$ derived for
neutron stars \cite{Ferrari2007} and have extrapolated their functional
dependence on compactness from the typical range relative to neutron
stars, \ie $\mu \in [0.10, 0.24]$, over to the typical values of
compactness that are relevant for gravastars, \ie $\mu \in [0.4,
  0.5)$. We note that the lower limit of this range of compactness is
  already rather small and that although gravastars with lower
  compactness can be constructed, they would not represent black-hole
  mimickers. Finally, although this represents a reasonable first
  approximation, especially since similar expressions have been shown to
  be valid for $\epsilon \lesssim 1$ \cite{Doneva:2013} and to be only
  mildly dependent on the equation of state \cite{Doneva:2013}, it is
  nevertheless an extrapolation and a perturbative analysis of rotating
  gravastars is needed to validate that the extrapolation is accurate.

\begin{figure}
\centering
\includegraphics[width=0.9\columnwidth]{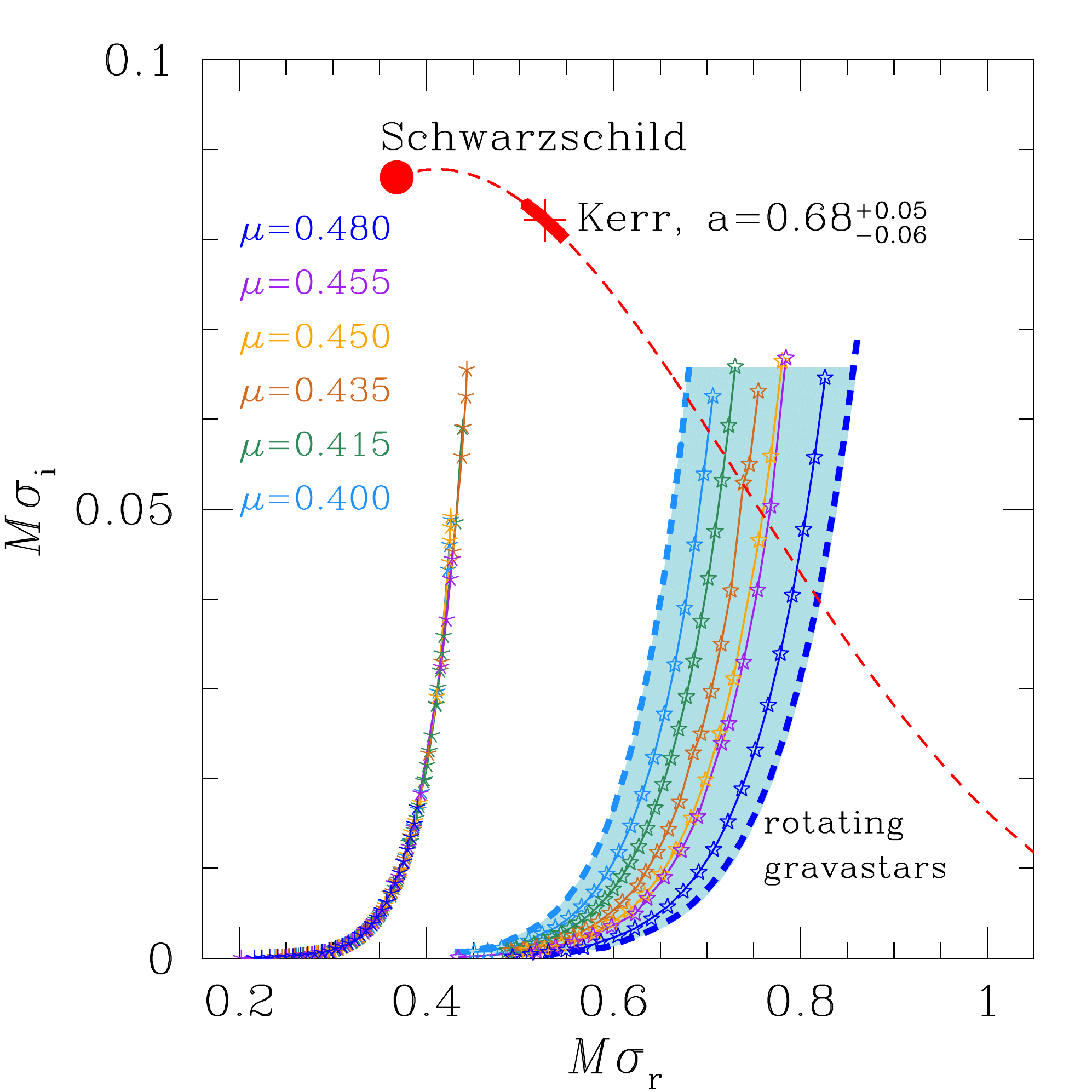}
\caption{The same as Fig. \ref{fig:f2} but including now also
  eigenfrequencies of rotating gravastars, as well as the
  eigenfrequencies of rotating black holes (red dashed line). Because of
  rotation the sequences of constant compactness no longer overlap and
  shaded area bounds the space of eigenfrequencies span by rotating
  gravastars compatible with GW150914. This region does not overlap by
  the range span by the eigenfrequencies of a Kerr black hole with
  dimensionless spin $a=0.68^{+0.05}_{-0.06}$ (thick red solid line).}
\label{fig:f3}
\end{figure}

We report in Fig. \ref{fig:f3} the real and imaginary parts of the
axial-mode eigenfrequencies relative to rotating gravastars as obtained
after computing the rotational and compactness corrections to the
eigenfrequencies for axial QNMs of spherical gravastars
\cite{chirenti_2007_htg}. In principle, polar modes would be the most
relevant for gravitational-wave emission since they excite fluid motions
\cite{Kokkotas99a}. However, for thin-shell gravastars, $\ell=2$ axial
and polar modes are almost isospectral for $\mu > 0.4$
\cite{Pani2009}. The solid lines of different colours in
Fig. \ref{fig:f3} represent sequences of constant-compactness rotating
gravastars with dimensionless spin $\chi=0.68$, which is a reasonable
prior given that gravastars are expected to have an orbital dynamics
similar to that of black holes (stable gravastars with large spin, $\chi
\lesssim 1.2$, are possible \cite{Chirenti2008}). Because gravastars will
have (slightly) larger sizes than black holes, the merger will
effectively take place a bit earlier in the inspiral (as it happens for
neutron stars) so that the effective final spin will be slightly larger
than the one assumed here \cite{Kastaun2013}. This is a rather
  crucial assumption, which is however rooted in the understanding built
  over the last 10 years when modelling the final spin from binary black
  holes in quasi-circular orbits; such a spin is ultimately determined by
  the combination of the initial spins of the black holes and of the
  angular momentum that is not radiated when the binary merges
  \cite{Barausse:2009uz, Healy2014, Hofmann2016}. As we discuss below,
  taking a value $\chi=0.68$ is actually a conservative choice.
  
Also in Fig. \ref{fig:f3}, symbols mark gravastars with different
thickness, which decreases when moving upwards along the curves. Note
that these sequences are no longer overlapping as rotation acts
differentially on gravastars of various compactness. In particular,
comparatively less compact gravastars will have smaller real
eigenfrequencies (light-blue solid line) than gravastars with larger
compactness (blue solid line). In addition, because the real part of the
eigenfrequencies increases with the rotation rate, the whole set of
curves moves to the right with $\epsilon$. Hence, the space of
eigenfrequencies spanned by rotating gravastars (shaded area) is effectively
bounded by the sequence with smallest compactness and rotation rate
within the measurements of GW150914, \ie $\mu = 0.40$ and
$\chi=0.68-0.06$ (light-blue thick dashed line), and by the sequence with
largest compactness and rotation rate, \ie $\mu = 0.48$ and
$\chi=0.68+0.05$ (dark-blue thick dashed line).

This space of eigenfrequencies between the two thick-dashed lines should
be compared with those relative to a rotating black hole and that is
marked with a dashed red line in Fig. \ref{fig:f3}, where the spin
obviously increases when moving from the left to the right along the
curve. Also marked with a square on the dashed line is the position of
the eigenfrequencies of a Kerr black hole with dimensionless spin
$a=0.68$. Finally, the red and thick solid portion of the dashed line
refers to eigenfrequencies with $a \in [0.68-0.06, 0.68+0.05]$
\cite{Abbott2016a,Abbott2016d}. Clearly, the range spanned by the
rotating-gravastars eigenfrequencies is distinct from the one spanned
when associating GW150914 to the ringdown of a rotating black
hole. Finally, considering a (slightly) larger value of $\chi$ on the
assumption that gravastars merge earlier than black holes, would just
move all the curves to the right, making the overlap even harder.

\begin{figure}
\centering
\includegraphics[width=0.9\columnwidth]{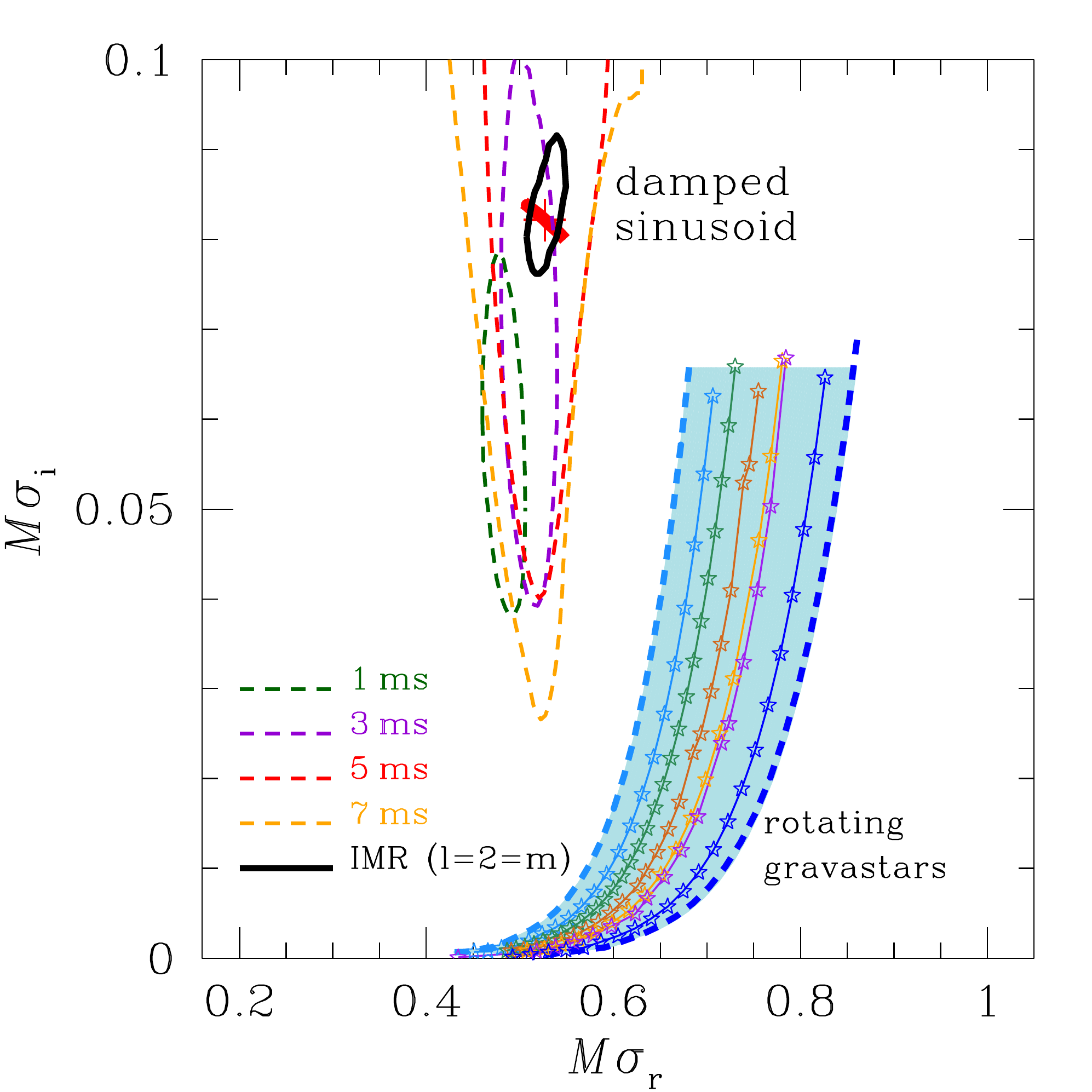}
\caption{The same as in Fig. \ref{fig:f3} but including also the 90\%
  confidence regions for the GW150914 ringdown frequencies, obtained with
  different starting times $t_0 = t_M + 1,3,5,7\ \textrm{ms}$ after the
  merger time $t_M$ \cite{Abbott2016c}. The solid black line shows the
  90\% confidence limit for the best fit combining the
  inspiral-merger-ringdown (IMR). Note that the Kerr eigenfrequencies are
  inside the IMR region and that there is no overlap between the contours
  and the gravastars eigenfrequencies.}
\label{fig:f4}
\end{figure}

In Fig. \ref{fig:f4} we compare our results to the GW150914 ringdown
frequencies that were obtained by a direct damped-sinusoid fit to the
ringdown data in \cite{Abbott2016c}. The different 90\% confidence
regions shown in Fig. \ref{fig:f4} correspond to different starting times
after the merger time, while the solid black line corresponds to the 90\%
confidence region obtained with the complete inspiral-merger-ringdown
(IMR) signal. In spite of the larger errors present, there is no overlap
between the GW150914 ringdown frequencies and the eigenfrequencies of
rotating gravastars, thus further strengthening our conclusions. These
results, notwithstanding the approximations employed here, lead us to the
conclusion that it is not possible to model the measured ringdown of
GW150914 as due to a rotating gravastar. 

Lastly, we should comment on the parameters of the final merged
object. We used in our analysis the values presented by the LIGO team
\cite{Abbott2016a,Abbott2016d}, which were obtained from an analysis of
the full GW150914 waveform (inspiral-merger-ringdown). If less
restrictive constraints on the parameters were to be considered, by
taking the 90\% confidence interval for the final mass and spin using
data only from the inspiral \cite{Abbott2016c}, our conclusions would
still hold. The larger uncertainties in the parameters only place a more
strict constraint on the compactness of the gravastar, but still well
within astrophysically relevant values (gravastars with $\mu > 0.48$
already show no overlap with the ringdown confidence regions shown in
Fig. \ref{fig:f4}).

\medskip\emph{Conclusions.~~} The measurement of the first direct
gravitational-wave signal GW150914 has provided the first strong evidence
for the existence of binary stellar-mass black hole systems. However, as
remarked in Ref. \cite{Abbott2016c}, it is not yet possible to set tight
constraints on different interpretations of GW150914 and, in particular,
on alternative theories to general relativity or on the idea that the
signal involves compact-object binaries composed of more exotic objects
such as boson stars \cite{Liebling2012} or gravastars
\cite{Mazur2004}. Hence, we have here considered the hypothesis that the
merging objects were indeed gravastars. Because the constraints coming
from GW150914 on the compactness of the merging objects are fully
compatible with the intrinsically large compactness that can be
associated with gravastars, it is presently difficult to exclude
gravastars as being responsible for the inspiral signal. In view of this,
we have concentrated on determining whether the merged object could be
interpreted as a rotating gravastar. To do this we have made use of the
numerous results available on the perturbations of nonrotating gravastars
and of rotating compact stars. In particular, we have modelled the
perturbative response of rotating gravastars as a correction to the
corresponding response of nonrotating gravastars; this approach is not
novel and it has been shown to be a successful route for computing the
eigenfrequencies of compact stars, either in slow rotation
\cite{Kojima1993, Ferrari2007} or in rapid rotation and within the
Cowling approximation \cite{Doneva:2013}. Using this approach and
comparing the real and imaginary parts of the ringdown signal with the
corresponding quantities for gravastars, we find that the range spanned
by the rotating-gravastars eigenfrequencies is well distinct from the one
spanned when associating the measured GW150914 signal to the ringdown of
a rotating black hole. Hence we conclude it is not possible to model the
measured ringdown of GW150914 as due to a rotating gravastar. While
  our analysis has considered gravastars of compactness $\mu \leq 0.48$,
  the behaviour of the QNM spectrum is such that our conclusions extend
  also to larger compactnesses.

We conclude with two final remarks. First, there is no contradiction
  between our results and those of Ref. \cite{Cardoso2016} as our
  conclusions refer to gravastars that are thick and have a surface at a
  small but not infinitesimal distance from the putative event
  horizon. Second, we stress again that our conclusion is based on a
technique that has been developed and employed robustly for rotating
compact stars and extended here to gravastars. This aspect of our
analysis calls for the development of a proper perturbative analysis of
rotating gravastars. Until such a framework is fully developed over the
coming years, our results can be used to provide an ``educated'' answer
to the question in the title.


\medskip\noindent\emph{Acknowledgements.~~} We thank E. Barausse,
V. Ferrari, K. Kokkotas, P. Pani, R. Sturani, and E. Berti for useful
discussions and comments. Support comes from the ERC Synergy Grant
``BlackHoleCam'' (Grant 610058), from ``NewCompStar'', COST Action
MP1304, from the LOEWE-Program in HIC for FAIR, from the European Union's
Horizon 2020 Research and Innovation Programme (Grant 671698) (call
FETHPC-1-2014, project ExaHyPE), and from the S\~ao Paulo Research
Foundation (FAPESP; Grant 2015/20433-4).


\bibliographystyle{apsrev4-1-noeprint}
\bibliography{aeireferences}


\end{document}